\documentclass{article}
\usepackage{spconf,amsmath,amssymb,graphicx,booktabs,multirow,hyperref}
\hypersetup{hidelinks}

\title{RMS-AQA: A Two-Stage Spatial Audio Question Answering  Benchmark for Real-World Domestic Environments}
\name{\begin{tabular}[t]{@{}c@{}}
      Peihao Chen$^1$, Qing Wang$^{1,*}$, Lichun Fan$^2$, Yufeng Hao$^3$, Zhifeng Kong$^4$, Mengyao Zhu$^5$, Hengyi Hong$^1$\\
      Hang Chen$^1$, Hang Su$^2$, Yujie Jian$^3$, Chao-Han Huck Yang$^{4,\dagger}$, Shichao Hu$^5$, Jun Du$^1$, Jian Luan$^2$, Ke Li$^3$\end{tabular}
      \thanks{$^*$corresponding author.}
      \thanks{$^{\dagger}$work done in NVIDIA.}}
\address{$^1$University of Science and Technology of China, China$\,$
         $^2$Independent Researcher, China\\
         $^3$DataoceanAI, China$\,$ $^4$NVIDIA, USA$\,$
         $^5$Soochow University, China}

\begin{document}
\ninept
\maketitle

\begin{abstract}
Embodied assistants in domestic environments must infer what happened, where and when it occurred, and how to respond. To address this, we introduce RMS-AQA, a spatial audio question answering (SAQA) benchmark for real-world domestic environments. The benchmark features a two-stage question-answering (QA) format to comprehensively assess the ability of audio-language models (ALMs) to first ground audible sound events and subsequently perform complex spatio-temporal reasoning based on that grounding. To maximize acoustic realism, our dataset combines authentic real-world first-order Ambisonics (FOA) recordings with high-fidelity synthetic data generated using measured room impulse responses (RIRs). Furthermore, we provide a lightweight spatial plug-in that injects FOA-format data into frozen audio-language backbones. Experimental results reveal that the primary challenges stem from concurrent sources, far distance, and sim-to-real domain gap between RIR-synthesized and authentic recordings. 
\end{abstract}


\section{Introduction}
\label{sec:intro}

An embodied assistant in a smart home must understand complex acoustic scenes rather than merely recognize isolated sounds. It must determine which events are present, where and when they occur, how concurrent events are related, and, when necessary, what response the situation warrants. This requirement necessitates the integration of acoustic perception with spatial and temporal reasoning. Consequently, an effective benchmark must measure whether a model can preserve information about grounded events while answering increasingly structured questions about the scene.

Recent large audio-language models (LALMs) have substantially improved general audio understanding. Open models such as Qwen3-Omni~\cite{Xu2025qwen3omni} support unified audio and audio-visual understanding; MiDashengLM~\cite{Dinkel2025midashenglm} uses general audio captions to cover speech, sounds, and music; and Audio Flamingo 3 (AF3)~\cite{Goel2025af3} and Audio Flamingo Next (AF-Next)~\cite{Ghosh2026afnext} extend open audio-language modeling to long recordings and temporally grounded reasoning. Closed models such as GPT-4o~\cite{OpenAI2024gpt4o} further demonstrate strong perception of speech style and emotion. However, these systems are predominantly trained and evaluated with monaural inputs. While they can describe what was heard, they lack direct access to the directional and distance cues needed to determine where a sound occurred or how sources are arranged within a room.

Recent work has begun to bridge this gap by extending audio question answering (AQA) to spatial signals. DCASE 2025 Task 5~\cite{Yang_2026} established a multi-domain setting for structured audio understanding, while BAT~\cite{Zheng2024learning} explored binaural question answering. Other studies introduce spatial branches or dedicated encoders for LALMs, including SPUR~\cite{Sakshi2025spur} and Spatial-Omni~\cite{Zhu2026spatial}. Meanwhile, SpatialSceneQA~\cite{You2026world} and STAR-Bench~\cite{Liu2025star} further probe spatial and spatio-temporal reasoning. These efforts demonstrate that spatial cues can improve localization and relational understanding. 

However, existing spatial audio question answering (SAQA) benchmarks suffer from a critical methodological limitation: they typically evaluate event recognition and spatio-temporal reasoning as isolated tasks, or they score only the final reasoning answer. This lack of explicit coupling creates a diagnostic blind spot. A model may output a correct reasoning answer based on entirely faulty event grounding (i.e., exploiting spurious correlations), or conversely, fail a reasoning task without the system being able to trace whether the error stemmed from acoustic perception or logical inference. 
From a data perspective, directly recorded domestic multichannel audio remains extremely scarce, and existing benchmarks heavily rely on purely simulated acoustic scenes. Most simulation pipelines utilize idealized, geometrically simplified room impulse responses (RIRs) that fail to capture the complex wave propagation and late reverberation. Consequently, models trained on these simplified simulations often fail to generalize to actual domestic environments. 

To address these limitations, we introduce RMS-AQA, a real-world multi-hop SAQA benchmark selected as an ICASSP 2027 Grand Challenge\footnotemark. RMS-AQA pairs first-order Ambisonics (FOA) audio clips with two-stage multiple-choice questions: Stage 1 grounds the audible sound events, and Stage 2 evaluates structured reasoning conditioned on that grounding. The model earns credit only when it answers both stages correctly, thereby certifying the complete grounding-to-reasoning chain while providing full diagnostic transparency. To support this task, we construct a large-scale, high-fidelity dataset of 177K clips. The training set is synthesized using measured real-room RIRs to maximize acoustic realism at scale, while the validation and test sets incorporate manually annotated recordings captured directly in real domestic environments. This dual-source composition allows us to evaluate the sim-to-real domain gap in spatial audio reasoning. We also provide a lightweight spatial plug-in that injects FOA-format data into a frozen LALM semantic backbone, enabling multichannel adaptation without discarding pretrained semantic knowledge.
\footnotetext{\textbf{\url{https://rmsaqachallenge.github.io/}}}

The main contributions are summarized as follows:
\begin{itemize}
    \item We introduce a two-stage SAQA task that explicitly couples event recognition with downstream scene reasoning, preventing lucky guesses and enabling precise error diagnosis.
    \item We construct a high-fidelity dataset that combines scalable training scenes synthesized with measured real-room RIRs and authentic real-world recordings for validation and test.
    \item We develop a plug-and-play spatial plug-in that augments a frozen LALM semantic encoder with FOA spatial cues for lightweight multichannel adaptation.
\end{itemize}
\section{Dataset and Benchmark}
\label{sec:rms-aqa}

\begin{figure}[t]
\centering
\includegraphics[width=\columnwidth]{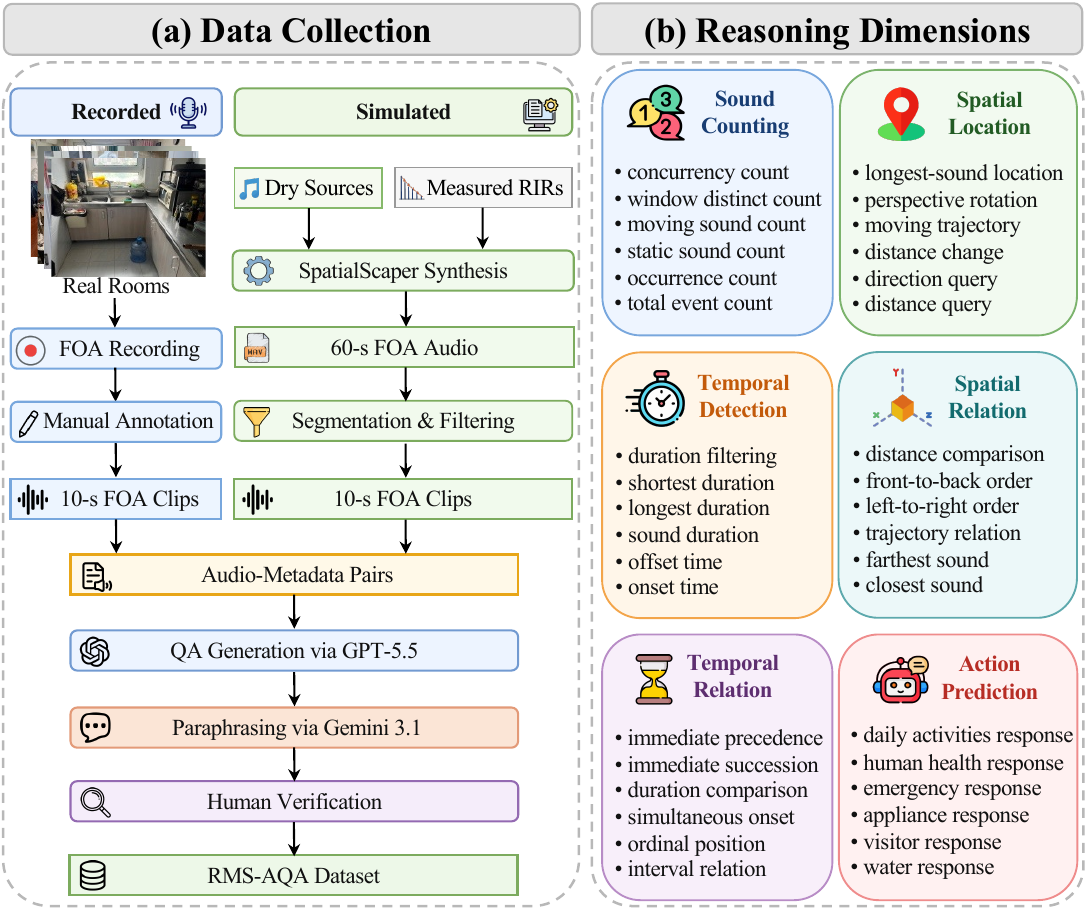}
\caption{Overview of the RMS-AQA benchmark. (a) The data collection pipeline. (b) The six reasoning dimensions of the benchmark.}
\label{fig:rms-aqa}
\end{figure}

\subsection{Data Collection}
\label{sec:data-collection}

As illustrated in Fig.~\ref{fig:rms-aqa}, the data collection pipeline of the RMS-AQA dataset encompasses both audio acquisition and QA generation. The audio data is constructed through a dual-track approach, combining authentic real-world recordings with high-fidelity simulated scenes. Both tracks ultimately yield 10-s, 24-kHz FOA clips paired with structured metadata.

\textbf{Recorded data track.} To capture the complex acoustics of actual domestic environments, we record four-channel FOA audio using a Zoom H3-VR device~\cite{zoomh3vr} across 18 distinct real rooms (e.g., living rooms, kitchens, bedrooms, and bathrooms). Specifically, 13 rooms are assigned to the validation set and 5 to the test set. The collection covers 49 sound classes. Participants perform scripted speech and household actions in stationary or moving configurations, forming scenes with up to four simultaneous sources. Common sounds are produced directly, while unsafe or hard-to-reproduce sounds are replayed through loudspeakers and re-recorded, preserving the authentic reverberation and spatial cues of the target rooms. Media sources utilize legally usable public-domain content, and specific sounds like infant cries include both live and replayed recordings. 
The continuous recordings are manually annotated with sound class, direction, height, distance, and temporal boundaries, and subsequently segmented into 10-s clips. Statistically, the recorded validation set exhibits a moving-source rate of 34.36\%, with a maximum observed source distance of 5.9 m.
\begin{table}[t]
\vspace{-0.1cm} 
\caption{Overview of the RMS-AQA dataset.}
\label{tab:split}
\centering
\setlength{\tabcolsep}{13pt}
\begin{tabular}{lccc}
\toprule
Dataset & \#\,Clips & \#\,QA pairs & \#\,Rooms \\
\midrule
Train      & 170K & 170K & 1,348 \\
Validation & 5K & 6K & 61 \\
Test       & 2K & 3K & 5 \\
Total      & 177K & 179K & 1,414 \\
\bottomrule
\end{tabular}
\vspace{-0.2cm} 
\end{table}

\textbf{Simulated data track.} We utilize SpatialScaper~\cite{roman2024spatialscaper} to synthesize high-fidelit scenes by convolving dry sound sources with measured FOA RIRs. The dry sources are drawn from AudioSet~\cite{gemmeke2017audioset}, FSD50K~\cite{fonseca2022fsd50k}, ESC-50~\cite{piczak2015esc50}, and CHiME-Home~\cite{Foster2015chimehome}, filtered to 324 domestic classes (e.g., speech, human activities, appliances, and pets). Measured RIRs are collected from FOA-MEIR~\cite{Yasuda2022foameir}, METU-SPARG~\cite{olgun2019metusparg}, TAU-SRIR~\cite{politis2022tausrir}, and ARNI-SRIR~\cite{McKenzie2021arni}, covering 1,396 rooms in total. To ensure rigorous evaluation, 1,348 rooms are allocated for training and 48 for simulated validation. Statistically, moving-source clips account for 45.04\% of the training set and 75.70\% of the simulated validation set. The maximum observed source distances reach 7.2 m for the training set and 6.4 m for the simulated validation set. The synthesized 60-s spatial audio is subsequently segmented and filtered into 10-s FOA clips.

\textbf{QA generation.} The audio clips from both tracks are accompanied by structured metadata capturing event identities, timestamps, directions, distances, and action-related risk indicators. Following the pipeline in Fig. 1(a), GPT-5.5~\cite{OpenAI2026gpt55} generates  two-stage, four-option questions from this metadata, and Gemini 3.1~\cite{Google2026gemini31pro} paraphrases them to enhance linguistic diversity. Human reviewers then verify each QA pair in the validation set, ensuring that the correct option is  unambiguous and consistent with the metadata, while the distractors remain plausible yet mutually distinct. The final RMS-AQA dataset allocates simulated clips for training, a mixture of simulated and recorded clips for validation, and exclusively recorded clips for testing. To prevent data leakage, the rooms used for RIR measurement in training and validation are mutually exclusive, as are the physical rooms used for recorded validation and testing. The detailed dataset split is summarized in Table~\ref{tab:split}.

\subsection{Task Protocol}
To explicitly evaluate the perception-to-reasoning chain, each FOA clip is paired with one or two two-stage QA pairs, with four answer options at each stage. Stage 1 first grounds the audible events. Stage 2 then receives the Stage-1 answer together with the audio and reasons about the grounded scene. This coarse-to-fine design gives Stage 2 an explicit event inventory before it tackles more complex questions. Crucially, it makes the source of an error observable: a failure may arise from incorrect sound recognition (Stage 1) or from flawed contextual reasoning (Stage 2). As detailed in Fig. 1(b), the benchmark evaluates models across six reasoning dimensions: sound counting (SC), spatial location (SL), temporal detection (TD), spatial relation (SR), temporal relation (TR), and action prediction (AP). These six categories are balanced across the benchmark, with the same average number of QA pairs per category.

\subsection{Evaluation Metric}
\label{sec:metric}

The primary metric of the benchmark is grounded Stage-2 accuracy, denoted ${\mathrm{A_{GS2}}}$:
\begin{equation}
{\mathrm{A_{GS2}}}=\frac{1}{N}\sum_{i=1}^{N}\mathbb{I}\!\left\{\hat{y}_i^{(1)}=y_i^{(1)}\land\hat{y}_i^{(2)}=y_i^{(2)}\right\}.
\end{equation}
Here, $y_i^{(1)}$ and $y_i^{(2)}$ denote the reference answers for the Stage-1 and Stage-2 questions of the $i$-th sample, respectively, while $\hat{y}_i^{(1)}$ and $\hat{y}_i^{(2)}$ represent their corresponding predictions. While Stage-1 accuracy evaluates event grounding, it provides limited insight into downstream reasoning. Conversely, raw Stage-2 accuracy is susceptible to inflation via lucky guesses, as a plausible answer might be credited even if the underlying event perception is wrong. By contrast, $\mathrm{A_{GS2}}$ assigns credit only when both stages are correct. This strict formulation not only prevents spurious guesses from yielding false positives, but also explicitly exposes how grounding errors propagate into subsequent reasoning. Stage-1 and raw Stage-2 accuracy are still reported to serve as complementary metrics. 

\begin{figure}[t]
\centering
\includegraphics[width=\columnwidth]{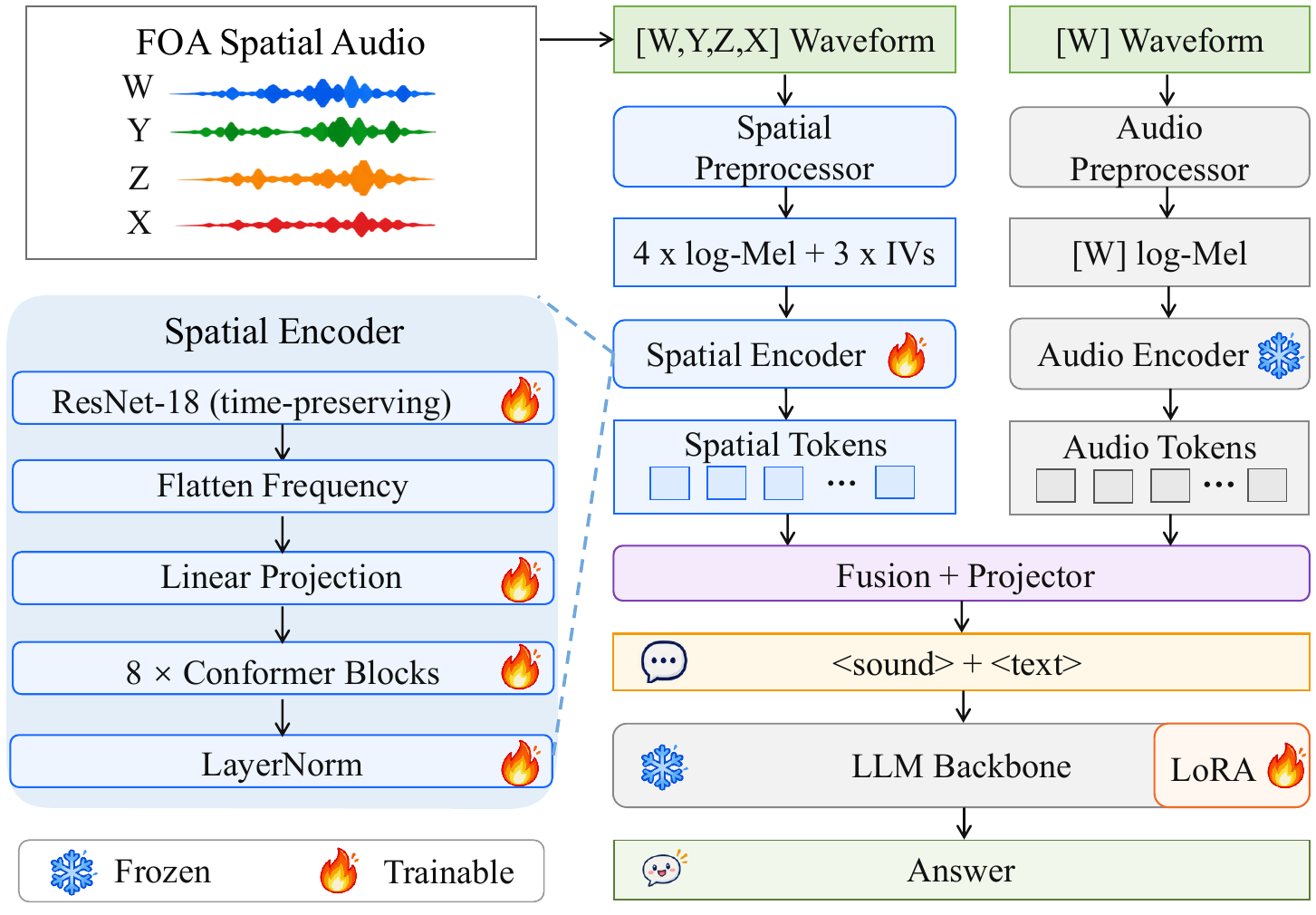}
\caption{Baseline overview. The frozen semantic branch and the trainable spatial branch are fused for two-stage answer generation.}
\label{fig:baseline}
\vspace{-0.2cm} 
\end{figure}

\begin{table*}[!t]
\caption{Overall and category-wise accuracy (\%) on the RMS-AQA validation set. $\mathrm{A_{S1}}$, $\mathrm{A_{S2}}$, and $\mathrm{A_{GS2}}$ denote Stage-1, raw Stage-2, and grounded Stage-2 accuracy, respectively. Rows marked ``+ SpatialAug'' report the adapted versions of the corresponding backbones.}
\label{tab:results}
\centering
%\footnotesize
\setlength{\tabcolsep}{3.5pt}
\renewcommand{\arraystretch}{1.12}
\begin{tabular}{@{}lccc@{\hspace{8pt}}cccccc@{\hspace{8pt}}cccccc@{}}
\toprule
\multirow{2}{*}{Model} & \multicolumn{3}{c}{Overall} & \multicolumn{6}{c}{Raw Stage-2 $\mathrm{A_{S2}}$} & \multicolumn{6}{c}{Grounded Stage-2 $\mathrm{A_{GS2}}$} \\
\cmidrule(lr){2-4}
\cmidrule(lr){5-10}
\cmidrule(lr){11-16}
 & $\mathrm{A_{S1}}$ & $\mathrm{A_{S2}}$ & $\mathrm{A_{GS2}}$ & SC & SL & TD & SR & TR & AP & SC & SL & TD & SR & TR & AP \\
\midrule
AF3~\cite{Goel2025af3} & 41.02 & 34.00 & 15.78 & 29.50 & 31.10 & 28.10 & 39.00 & 40.10 & 36.20 & 9.80 & 14.60 & 14.60 & 17.30 & 20.80 & 17.60 \\
+\,SpatialAug & 63.98 & 52.60 & 39.10 & 59.10 & 36.40 & 48.90 & 45.40 & 57.30 & 68.50 & 46.40 & 25.20 & 36.90 & 29.80 & 44.30 & 52.00 \\
\hline
AF-Next~\cite{Ghosh2026afnext}& 41.18 & 35.57 & 15.95 & 35.40 & 31.90 & 28.30 & 37.70 & 40.30 & 39.80 & 13.10 & 14.20 & 13.20 & 16.10 & 19.90 & 19.20 \\
+\,SpatialAug & 65.93 & 60.23 & 45.17 & 69.30 & 40.80 & 62.70 & 47.90 & 62.20 & 78.50 & \textbf{54.50} & 29.00 & 46.40 & 33.40 & 49.30 & 58.40 \\
\hline
MiDashengLM~\cite{Dinkel2025midashenglm} & 23.63 & 34.97 & 9.40 & 46.20 & 28.40 & 24.00 & 36.40 & 39.60 & 35.20 & 13.90 & 6.80 & 6.70 & 9.60 & 10.80 & 8.60 \\
+\,SpatialAug & \textbf{66.85} & 59.70 & 44.68 & 66.10 & 40.20 & 62.80 & 47.10 & 60.70 & \textbf{81.30} & 50.00 & 28.80 & 48.10 & 33.40 & 47.70 & 60.10 \\
\hline
Qwen3-Omni~\cite{Xu2025qwen3omni} & 37.70& 34.90 & 16.72 & 43.40 & 26.80 & 32.60 & 27.80 & 40.70 & 38.10 & 17.90 & 13.30 & 15.50 & 14.40 & 22.40 & 16.80 \\
+\,SpatialAug & 66.70 & \textbf{63.45} & \textbf{48.93} & \textbf{69.40} & \textbf{42.20} & \textbf{71.20} & \textbf{52.50} & \textbf{66.40} & 79.00 & 54.20 & \textbf{31.40} & \textbf{54.40} & \textbf{38.30} & \textbf{54.40} & \textbf{60.90} \\
\midrule
GPT-Audio-1.5~\cite{OpenAI2026gptaudio15} & 37.28 & 33.05 & 17.07 & 43.50 & 21.70 & 25.70 & 21.80 & 37.30 & 48.30 & 23.60 & 10.80 & 12.40 &11.50 & 19.30 & 24.80\\
Gemini-3.1-Pro~\cite{Google2026gemini31pro} & 52.45 & 46.42 & 29.55 & 50.00 & 29.40 & 56.40 & 33.40 & 57.20 & 52.10 & 32.70 & 18.30 & 32.50 & 22.90 & 36.80 & 34.10 \\
Qwen3.8-Omni-Flash~\cite{qwen2026omniflash}&59.63 & 54.53& 36.38&57.80 &36.60 &68.50 &38.40 &64.30 &61.60 &39.40 &24.40 &44.10 &26.80 & 43.60&40.00\\
\bottomrule
\end{tabular}
\vspace{-0.2cm} 
\end{table*}

\section{Baseline}
\label{sec:baseline-method}

\subsection{Spatial-Augmented Architecture}
\label{sec:baseline}

As illustrated in Fig.~\ref{fig:baseline}, our baseline augments a monaural audio encoder with a trainable spatial branch. To preserve the pretrained semantic knowledge, the audio encoder and the large language model (LLM) weights are kept strictly frozen. The audio encoder extracts semantic tokens solely from the omnidirectional FOA $W$ channel. In parallel, the spatial plug-in receives the full four-channel FOA representation and supplies directional cues through a temporally aligned encoder. This decoupled design effectively exposes spatial information without disrupting the monaural semantic backbone. 

For the semantic branch, log-Mel features extracted from the FOA $W$ channel are fed into the frozen audio encoder, yielding the semantic representation $\mathrm{\mathbf{H}^{sem}}\in\mathbb{R}^{T\times D_{\mathrm{sem}}}$, where $T$ denotes the number of temporal tokens and $D_{\mathrm{sem}}$ is the semantic feature dimension. For the spatial branch, following our previous works~\cite{wang2023four,dong2025experimental}, the 4-channel log-Mel features and the 3-channel intensity vectors (IVs) are concatenated as input. Thus, the spatial encoder processes seven aligned feature maps, whereas the audio encoder uses only the $W$-channel input. Within the spatial encoder, a time-preserving ResNet-18~\cite{he2016deep} first aggregates local time--frequency patterns by pooling only along the frequency axis. After frequency flattening, a linear projection expands the token dimension to 1024. Subsequently, eight Conformer blocks model longer-range temporal dependencies through self-attention and convolutional mixing, followed by layer normalization, yielding $\mathrm{\mathbf{H}^{spa}}\in\mathbb{R}^{T\times1024}$.

The fusion module is designed to be lightweight. Let $\operatorname{Concat}(\cdot,\cdot)$ denote the concatenation operator. The semantic and spatial tokens are concatenated and projected to the LLM dimension $D_{\mathrm{llm}}$:
\begin{equation}
\mathrm{\mathbf{H}^{fuse}}=\mathbf{W}\operatorname{Concat}\!\left(\mathrm{\mathbf{H}^{sem}},\mathrm{\mathbf{H}^{spa}}\right)+\mathbf{b}.
\end{equation}
where $\mathrm{\mathbf{H}^{fuse}}\in\mathbb{R}^{T\times D_{\mathrm{llm}}}$ represents the fused audio-spatial tokens fed into the LLM backbone. Crucially, the projection weight $\mathbf{W}$ and its bias $\mathbf{b}$ are zero-initialized, ensuring that the adaptation smoothly starts from the semantic-only mapping. 

\subsection{Training Details}
\label{sec:training-details}

As indicated in Fig.~\ref{fig:baseline}, the audio encoder and base LLM parameters remain frozen; only the spatial encoder, the fusion projection, and the rank-$r$ low-rank adaptation (LoRA) adapters \cite{hu2022lora} on the LLM receive gradient updates. Both QA stages utilize autoregressive next-token supervision. During training, Stage 2 is conditioned on the reference Stage-1 answer to form the conversational context. Let $\mathcal{L}_1$ and $\mathcal{L}_2$ be the mean next-token cross-entropies over $N_1$ and $N_2$ non-padding answer tokens, respectively. The joint loss is formulated as:
\begin{equation}
\mathcal{L}=\frac{N_1\mathcal{L}_1+N_2\mathcal{L}_2}{N_1+N_2}.
\end{equation}

The model is trained for three epochs using the AdamW optimizer with a learning rate of $2\times10^{-4}$, a weight decay of $10^{-2}$, and global gradient-norm clipping at 5.0. Distributed data parallelism is employed across four NVIDIA A100 GPUs, with a per-GPU batch size of 8. The baseline is adapted using rank-16 LoRA ($\alpha=16$, dropout 0.05) across all 28 LLM layers.

During inference, the model strictly follows the deployed two-stage path: Stage 1 is decoded greedily for a maximum of 48 new tokens, and the predicted Stage-1 answer is inserted into the Stage-2 prompt. This preserves the intended progression from event grounding to higher-order reasoning. For each run, the checkpoint achieving the highest $\mathrm{A_{GS2}}$ on the validation set is selected for evaluation.

\section{Results and Analysis}
\label{sec:experiments}

\subsection{Evaluated Systems}

We evaluate our proposed spatial adaptation method on several open-weight ALMs, including AF3~\cite{Goel2025af3}, AF-Next~\cite{Ghosh2026afnext}, MiDashengLM~\cite{Dinkel2025midashenglm}, and Qwen3-Omni~\cite{Xu2025qwen3omni}. For these models, we compare their monaural zero-shot performance against their spatial-augmented versions. Additionally, GPT-Audio-1.5~\cite{OpenAI2026gptaudio15} Gemini-3.1-Pro~\cite{Google2026gemini31pro}, and Qwen3.8-Omni-Flash~\cite{qwen2026omniflash} serve as closed-source zero-shot baselines. Among them, Qwen3.8-Omni-Flash is particularly notable as it natively supports FOA input.

\subsection{Results and Analysis}

Table~\ref{tab:results} reports overall and category-wise performance on the validation set. Spatial augmentation improves all three overall metrics across the four adapted backbones, with gains in grounded Stage-2 accuracy ranging from 23.32\% to 35.28\% over their monaural zero-shot counterparts. The adapted Qwen3-Omni achieves the highest overall grounded accuracy of 48.93\%, improving over its zero-shot counterpart by 32.21\%. It also outperforms Qwen3.8-Omni-Flash, which supports native FOA input and is the strongest closed-source baseline, by 12.55\%. Other closed-source models, such as Gemini-3.1-Pro, trail behind both Qwen3.8-Omni-Flash and all four spatial-augmented systems.

Across categories, spatial augmentation yields the most pronounced gains in SC and AP, as improved event grounding inherently enhances source separation for counting and provides richer semantic context for decision-making. The category-wise results reveal persistent challenges in spatial understanding. Although spatial augmentation improves grounded accuracy in every category for all four backbones, SL and SR remain the two lowest-scoring categories for each adapted model. Even the strongest system achieves only 31.40\% and 38.30\% grounded accuracy on SL and SR, respectively, indicating substantial room for improvement in spatial localization and relational reasoning. AP achieves the best performance among all categories, demonstrating that reliable sound identification is crucial for risk-aware embodied decision-making. Moreover, its raw Stage-2 accuracy of 63.45\% falls to 48.93\% when correct Stage-1 grounding is also required. This 14.52\% gap shows that correct downstream answers do not necessarily coincide with correct event grounding, highlighting the necessity of our strict $\mathrm{A_{GS2}}$ metric to penalize spurious guesses and false positives.

\begin{table}[t]
\caption{Ablation on Stage-1 context for raw Stage-2 accuracy (\%). ``w/o Context'' evaluates context-free reasoning. ``Predicted'' and ``Reference'' use generated and oracle Stage-1 answers, respectively.}
\label{tab:two-stage}
\centering
\small
\setlength{\tabcolsep}{4pt}
\renewcommand{\arraystretch}{1.15}
\begin{tabular}{lccc}
\toprule
Model & w/o Context & Predicted & Reference \\
\midrule
AF3 + SpatialAug & 44.12 & 52.60 & \textbf{60.05} \\
Qwen3-Omni + SpatialAug & 58.00 & 63.45 & \textbf{70.40} \\
\bottomrule
\end{tabular}
\vspace{-0.2cm} 
\end{table}

Table 3 evaluates the necessity of Stage-1 contextual conditioning for Stage-2 reasoning. Across both backbone architectures, explicitly providing Stage-1 event grounding greatly boosts Stage-2 reasoning accuracy. Specifically, answering Stage 2 conditioned on the predicted Stage-1 answer yields substantial absolute gains over the context-free baseline (+8.48\% for AF3 and +5.45\% for Qwen3-Omni), reaching 52.60\% and 63.45\%, respectively. Conditioning on the oracle reference Stage-1 answer further unlocks the reasoning potential, pushing accuracies to 60.05\% (+15.93\%) and 70.40\% (+12.40\%). This progressive improvement demonstrates that explicit event grounding provides essential perceptual priors for downstream reasoning. Furthermore, the about 7\% gap between predicted and reference highlights that grounding errors remain the primary bottleneck, demonstrating how our two-stage design isolates perception failures from reasoning deficits.

\begin{figure}[!t]
\centering
\includegraphics[width=\columnwidth]{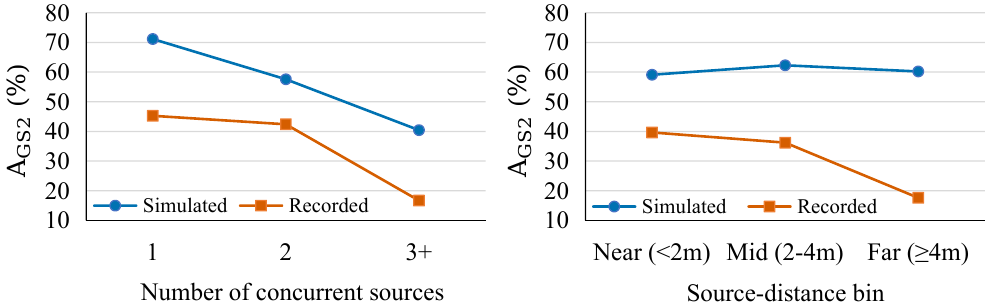}
\caption{Impact of scene complexity on grounded Stage-2 accuracy for the adapted Qwen3-Omni across simulated and recorded data.}
\label{fig:difficulty}
\vspace{-0.2cm} 
\end{figure}

Fig.~\ref{fig:difficulty} illustrates how scene complexity and the sim-to-real domain gap jointly affect reasoning performance. Real recordings prove significantly more challenging, evidenced by a 23.67\% performance drop for the adapted Qwen3-Omni compared to simulated scenes. As concurrent sources increase, accuracy degrades steadily due to severe time-frequency interference and overlapping spatial cues. More notably, while simulated performance remains robust across distances, accuracy on real recordings drops sharply for far-field sources ($\geq$4m). This sharp decline underscores the acoustic mismatch between RIR-based scene synthesis and real-world capture, where complex reverberation and weak direct sounds severely degrade spatial reliability. 

\section{Conclusion and limitations}
This paper presented RMS-AQA, a comprehensive benchmark for spatial audio reasoning in domestic environments. Our two-stage evaluation protocol successfully isolates perception failures from reasoning deficits, preventing score inflation from spurious guesses. Additionally, we demonstrated that a lightweight spatial plug-in can effectively inject multichannel cues into frozen audio-language backbones, driving significant performance improvements. Evaluations across our high-fidelity simulated and authentic recorded data sets highlight that overlapping sources, far-field acoustics, and the sim-to-real domain gap remain critical bottlenecks. Future research will aim to bridge this acoustic mismatch and further improve fine-grained spatial localization in complex real-world scenarios.

%\clearpage
%\begingroup
%\def\section#1{\par\begingroup\centering\bfseries\MakeUppercase{#1}\par\endgroup\vspace{0.8ex}}
\bibliographystyle{IEEEbib}
\bibliography{refs}

@inproceedings{roman2024spatialscaper,
  title={Spatial scaper: a library to simulate and augment soundscapes for sound event localization and detection in realistic rooms},
  author={Roman, Iran R and Ick, Christopher and Ding, Sivan and Roman, Adrian S and McFee, Brian and Bello, Juan P},
  booktitle={IEEE International Conference on Acoustics, Speech and Signal Processing},
  pages={1221--1225},
  year={2024}
}

@article{olgun2019metusparg,
  title={{METU SPARG} eigenmike em32 acoustic impulse response dataset v0. 1.0},
  author={Olgun, Orhun and Hacihabiboglu, Huseyin},
  journal={Graduate School Inform., Middle East Tech. Univ., Ankara, Turkey, Tech. Rep},
  pages={5},
  year={2019}
}

@article{politis2022tausrir,
  title={{TAU spatial room impulse response database (TAU-SRIR DB)}},
  author={Politis, Archontis and Adavanne, Sharath and Virtanen, Tuomas},
  journal={Zenodo},
  year={2022}
}

@inproceedings{Yasuda2022foameir,
  title={Echo-aware adaptation of sound event localization and detection in unknown environments},
  author={Yasuda, Masahiro and Ohishi, Yasunori and Saito, Shoichiro},
  booktitle={IEEE International Conference on Acoustics, Speech and Signal Processing},
  pages={226--230},
  year={2022}
}

@article{McKenzie2021arni,
  author  = {McKenzie, Thomas and McCormack, Leo and Hold, Christoph},
  title   = {Dataset of Spatial Room Impulse Responses in a Variable Acoustics Room for Six Degrees-of-Freedom Rendering and Analysis},
  journal = {arXiv preprint arXiv:2111.11882},
  year    = {2021},
  url     = {https://arxiv.org/abs/2111.11882}
}

@inproceedings{gemmeke2017audioset,
  title={Audio set: An ontology and human-labeled dataset for audio events},
  author={Gemmeke, Jort F and others},
  booktitle={IEEE International Conference on Acoustics, Speech and Signal Processing},
  pages={776--780},
  year={2017}
}

@article{fonseca2022fsd50k,
  title={{FSD50K}: an open dataset of human-labeled sound events},
  author={Fonseca, Eduardo and Favory, Xavier and Pons, Jordi and Font, Frederic and Serra, Xavier},
  journal={IEEE/ACM Transactions on Audio, Speech, and Language Processing},
  volume={30},
  pages={829--852},
  year={2021},
  publisher={IEEE}
}

@inproceedings{piczak2015esc50,
  title={{ESC}: Dataset for environmental sound classification},
  author={Piczak, Karol J},
  booktitle={ACM International Conference on Multimedia},
  pages={1015--1018},
  year={2015}
}

@misc{zoomh3vr,
  author       = {{Ambisonic Studio}},
  title        = {{Zoom H3-VR -- Ambisonic Recorder}},
  year         = {2024},
  howpublished = {\url{https://ambisonic.studio/zoom-h3-vr/index.html}}
}

@article{Goel2025af3,
  title={Audio flamingo 3: Advancing audio intelligence with fully open large audio language models},
  author={Ghosh, Sreyan and Goel, Arushi and Kim, Jaehyeon and others},
  journal={Advances in Neural Information Processing Systems},
  volume={38},
  pages={41819--41886},
  year={2026}
}

@article{Xu2025qwen3omni,
  title={Qwen3-omni technical report},
  author={Xu, Jin and Guo, Zhifang and Hu, Hangrui and others},
  journal={arXiv preprint arXiv:2509.17765},
  year={2025}
}

@article{Dinkel2025midashenglm,
  title={{MiDashengLM}: Efficient audio understanding with general audio captions},
  author={Dinkel, Heinrich and Li, Gang and Liu, Jizhong and others},
  journal={arXiv preprint arXiv:2508.03983},
  year={2025}
}

@article{Ghosh2026afnext,
  author       = {Ghosh, Sreyan and Goel, Arushi and Jayakumar, Kaousheik and others},
  title        = {Audio flamingo next: Next-generation open audio-language models for speech, sound, and music},
  journal = {arXiv preprint arXiv:2604.10905},
  year         = {2026}
}

@article{OpenAI2024gpt4o,
  title={{GPT}-4o system card},
  author={Hurst, Aaron and Lerer, Adam and Goucher, Adam P and others},
  journal={arXiv preprint arXiv:2410.21276},
  year={2024}
}

@misc{OpenAI2026gpt55,
  author       = {{OpenAI}},
  title        = {{GPT-5.5 system card}},
  year         = {2026},
  howpublished = {\url{https://openai.com/index/gpt-5-5-system-card/}}
}

@misc{OpenAI2026gptaudio15,
  author       = {{OpenAI}},
  title        = {{GPT-Audio-1.5}},
  year         = {2026},
  howpublished = {\url{https://developers.openai.com/api/docs/models/gpt-audio-1.5}}
}

@misc{Google2026gemini31pro,
  author       = {{Google DeepMind}},
  title        = {{Gemini 3.1 Pro model card}},
  year         = {2026},
  howpublished = {\url{https://deepmind.google/models/model-cards/gemini-3-1-pro/}}
}

@inproceedings{Yang_2026,
  author    = {Yang, Chao-Han Huck and Ghosh, Sreyan and Wang, Qing and others},
  title     = {Multi-domain audio question answering benchmark toward acoustic content reasoning},
  booktitle = {IEEE International Conference on Acoustics, Speech and Signal Processing},
  year      = {2026},
  pages     = {22517--22521},
  doi       = {10.1109/ICASSP55912.2026.11464798}
}

@inproceedings{Zheng2024learning,
author = {Zheng, Zhisheng and Peng, Puyuan and Ma, Ziyang and Chen, Xie and Choi, Eunsol and Harwath, David},
title = {{BAT}: Learning to reason about spatial sounds with large language models},
year = {2024},
pages = {61454--61469},
booktitle = {International Conference on Machine Learning}
}

@article{Sakshi2025spur,
  title={{SPUR}: A plug-and-play framework for integrating spatial audio understanding and reasoning into large audio-language models},
  author={Sakshi, S and Lokegaonkar, Vaibhavi and Zhang, Neil and others},
  journal={arXiv preprint arXiv:2511.06606},
  year={2025}
}

@article{Zhu2026spatial,
  title={{Spatial-Omni}: Spatial audio understanding integration in multimodal LLMs via FOA encoding},
  author={Zhu, Zhiyuan and Chen, Yixuan and Shao, Yiwen and others},
  journal={arXiv preprint arXiv:2606.10738},
  year={2026}
}

@article{You2026world,
  title={The world is not mono: Enabling spatial understanding in large audio-language models},
  author={You, Yuhuan and Wei, Lai and Wu, Xihong and Qu, Tianshu},
  journal={arXiv preprint arXiv:2601.02954},
  year={2026}
}

@inproceedings{Liu2025star,
  title={{STAR-Bench}: Probing deep spatio-temporal reasoning as audio 4d intelligence},
  author={Liu, Zihan and Niu, Zhikang and Xiao, Qiuyang and others},
  booktitle={International Conference on Learning Representations},
  volume={2026},
  pages={134703--134731},
  year={2026}
}

@inproceedings{he2016deep,
  author    = {He, Kaiming and Zhang, Xiangyu and Ren, Shaoqing and Sun, Jian},
  title     = {Deep Residual Learning for Image Recognition},
  booktitle = {IEEE Conference on Computer Vision and Pattern Recognition},
  pages     = {770--778},
  year      = {2016},
  doi       = {10.1109/CVPR.2016.90}
}

@inproceedings{hu2022lora,
  author    = {Hu, Edward J. and Shen, Yelong and Wallis, Phillip and others},
  title     = {{LoRA}: Low-Rank Adaptation of Large Language Models},
  booktitle = {International Conference on Learning Representations},
  year      = {2022},
  url       = {https://openreview.net/forum?id=nZeVKeeFYf9}
}

@inproceedings{Foster2015chimehome,
  author    = {Foster, Peter and Sigtia, Siddharth and Krstulovic, Sacha and Barker, Jon and Plumbley, Mark D.},
  title     = {{CHiME-Home}: A Dataset for Sound Source Recognition in a Domestic Environment},
  booktitle = {IEEE Workshop on Applications of Signal Processing to Audio and Acoustics},
  year      = {2015},
  doi       = {10.1109/WASPAA.2015.7336899}
}

@misc{qwen2026omniflash,
  author       = {{Qwen Team}},
  title        = {{Qwen3.8-Omni-Flash}: Omni Senses. Agentic Delivery},
  year         = {2026},
  howpublished = {\url{https://www.qianwenai.com/models/qwen3.8-omni-flash}}
}

@article{wang2023four,
  title={A four-stage data augmentation approach to resnet-conformer based acoustic modeling for sound event localization and detection},
  author={Wang, Qing and Du, Jun and Wu, Hua-Xin and Pan, Jia and Ma, Feng and Lee, Chin-Hui},
  journal={IEEE/ACM Transactions on Audio, Speech, and Language Processing},
  volume={31},
  pages={1251--1264},
  year={2023},
  publisher={IEEE}
}

@inproceedings{dong2025experimental,
  title={An experimental study on joint modeling for sound event localization and detection with source distance estimation},
  author={Dong, Yuxuan and Wang, Qing and Hong, Hengyi and Jiang, Ya and Cheng, Shi},
  booktitle={IEEE International Conference on Acoustics, Speech and Signal Processing},
  pages={1--5},
  year={2025}
}
%\endgroup

\end{document}